\documentclass[numbers]{elsarticle}
\usepackage{latexsym}
\usepackage{natbib}
\usepackage[latin2]{inputenc}
\usepackage{longtable}
\RequirePackage{lineno} 

\makeatletter
\def\ps@pprintTitle{%
 \let\@oddhead\@empty
 \let\@evenhead\@empty
 \def\@oddfoot{\centerline{\thepage}}%
 \let\@evenfoot\@oddfoot}
\makeatother

\begin{document}

\title{Graph Theoretical Analysis Reveals: Women's Brains are Better Connected than Men's}

\author[p]{Balázs Szalkai}
\ead{szalkai@pitgroup.org}
\author[p]{Bálint Varga}
\ead{varga@pitgroup.org}
\author[p,u]{Vince Grolmusz\corref{cor1}}
\ead{grolmusz@pitgroup.org}
\cortext[cor1]{Corresponding author}
\address[p]{PIT Bioinformatics Group, Eötvös University, H-1117 Budapest, Hungary}
\address[u]{Uratim Ltd., H-1118 Budapest, Hungary}

\date{}

\begin{abstract}
Deep graph-theoretic ideas in the context with the graph of the World Wide Web led to the definition of Google's PageRank and the subsequent rise of the most-popular search engine to date. 
Brain graphs, or connectomes, are being widely explored today. We believe that non-trivial graph theoretic concepts, similarly as it happened in the case of the World Wide web, will lead to discoveries enlightening the structural and also the functional details of the animal and human brains.	
When scientists examine large networks of tens or hundreds of millions of vertices, only fast algorithms can be applied because of the size constraints. In the case of diffusion MRI-based structural human brain imaging, the effective vertex number of the connectomes, or brain graphs derived from the data is on the scale of several hundred today. That size facilitates applying strict mathematical graph algorithms even for some hard-to-compute (or NP-hard) quantities like vertex cover or balanced minimum cut. 

In the present work we have examined brain graphs, computed from the data of the Human Connectome Project, recorded from male and female subjects between ages 22 and 35. Significant differences were found between the male and female structural brain graphs: we show that the average female connectome has more edges, is a better expander graph, has larger minimal bisection width, and has more spanning trees than the average male connectome. Since the average female brain weights less than the brain of males, these properties show that the female brain is more ``well-connected'' or perhaps, more ``efficient'' in a sense than the brain of males.  It is known that the female brain has a larger white matter/gray matter ratio than the brain of males; this observation is in line with our findings concerning the number of edges, since  the white matter consists of myelinated axons, which, in turn, correspond to the connections in the brain graph. We have also found that the minimum bisection width, normalized with the edge number, is also significantly larger in the right and the left hemispheres in females: therefore, that structural difference is independent from the difference in the number of edges. 
\end{abstract}

\maketitle

\section{Introduction}

In the last several years hundreds of publications appeared describing or analyzing structural or functional networks of the brain, frequently referred to as "connectome" \cite{Hagmann2012,Craddock2013a}. Some of these publications analyzed data from healthy humans \cite{Ball2014,Bargmann2012,Batalle2013,Graham2014}, and some compared the connectome of the healthy brain with diseased one \cite{Agosta2014,AlexanderBloch2014,Baker2014,Besson2014a,Bonilha2014}. 

So far, the analyses of the connectomes mostly used tools developed for very large networks, such as the graph of the World Wide Web (with billions of vertices), or protein-protein interaction networks (with tens or hundreds of thousands of vertices), and because of the huge size of original networks, these methods used only very fast algorithms and frequently just primary degree statistics and graph-edge counting between pre-defined regions or lobes of the brain \cite{Ingalhalikar2014b}. 

In the present work we demonstrate that deep and more intricate graph theoretic parameters could also be computed by using, among other tools, contemporary integer programming approaches for connectomes with several hundred vertices. 

With these mathematical tools we show statistically significant differences in some graph properties of the connectomes, computed from MRI imaging data of male and female brains. We will not try to associate behavioral patterns of males and females with the discovered structural differences \cite{Ingalhalikar2014b} (see also the debate that article has generated: \cite{Joel2014,Ingalhalikar2014a,Fine2014}), because we do not have behavioral data of the subjects of the imaging study, and, additionally, we cannot describe high-level functional properties implied by those structural differences. However, we clearly demonstrate that deep graph-theoretic parameters show "better" connections in a certain sense in female connectomes than in male ones.

The study of \cite{Ingalhalikar2014b} analyzed the 95-vertex graphs of 949 subjects aged between 8 and 22 years, using basic statistics for the numbers of edges running either between or within different lobes of the brain (the parameters deduced were called {\em hemispheric connectivity ratio, modularity, transitivity and participation coefficients}, see \cite{Ingalhalikar2014b} for the definitions). It was found that males have significantly more intra-hemispheric edges than females, while females have significantly more inter-hemispheric edges than males.

\section{Results and Discussion}

We have analyzed the connectomes of 96 subjects, 52 females and 44 males, each with 83, 129 and 234 node resolutions, and each graphs with five different weight functions. We considered the connectomes as graphs with weighted edges, and performed graph-theoretic analyses with computing some polynomial-time computable and also some NP-hard graph parameters on the individual graphs, and then compared the results statistically for the male and the female group. 

We have found that female connectomes have more edges, larger (normalized) minimum bisection widths, larger minimum-vertex covers and more spanning trees than the male connectomes. 

In order to describe the parameters, which differ significantly among male and female connectomes, we need to place them in the context of their graph theoretical definitions.

\subsection{Edge number and edge weights}

We have found significantly higher number of edges (counted with 5 types of weights and also without any weights) in both hemispheres and also in the whole brain in females, in all resolutions. This finding is surprising, since we used the same parcellation and the same tractography and the same graph-construction methods for female and male brains, and because it is proven that females have, on average, less-weighting brains than males \cite{Witelson2006}. For example, in the 234-vertex resolution, the average number of (unweighted) edges in female connectomes is $1826$, in males $1742$, with $p=0.00063$ (see the Appendix for tables with the results). The work of \cite{Ingalhalikar2014b} reported similar findings in inter-hemispheric connections only.

It is known that there are statistical differences in the size and the weight of the female and the male cerebra \cite{Witelson2006}. 
It was also published \cite{Taki2011} that female brains statistically have a higher white matter/gray matter ratio than male brains. We argue that this observation is in line with the quantitative differences in the fibers and edges in the connectomes of the sexes.

In a simplified view, the edges of the braingraph correspond to the fibers of the myelinated axons in the white matter, while the nodes of the graph to areas of the gray matter. Therefore, since females have a higher white matter/gray matter ratio than males by \cite{Taki2011} that fact implies that the number of detected fibers by the tractography step of the processing is relatively higher in females than in males, and this higher number of fibers imply higher number of edges in female connectomes.

\subsection{Minimum cut and balanced minimum cut} 

Suppose the nodes, or the vertices, of a graph are partitioned into two, disjoint, non-empty sets, say $X$ and $Y$; their union is the whole vertex-set of the graph. The $X,Y$ {\em cut} is the set of {\em all} edges connecting vertices of $X$ with the vertices of $Y$ (Figure 1A). The size of the cut is the number of edges in the cut. In graph theory, the size of the minimum cut is an interesting quantity. The minimum cut between vertices $a$ and $b$ is the minimum cut, taken for all $X$ and $Y$, where vertex $a$ is in $X$ and $b$ is in $Y$. This quantity gives the ``bottleneck'', in a sense, between those two nodes (c.f., Menger theorems and Ford-Fulkerson's Min-Cut-Max-Flow theorem \cite{Lawler1976, Ford1956}). The minimum cut in a graph is defined to be the cut with the fewest edges for {\em all} non-empty sets $X$ and $Y$, partitioning the vertices.  

Clearly, for non-negative weights, the size of the minimum cut in a non-connected graph is 0. Very frequently, however, in connected graphs, the minimum cut is determined by just the smallest degree node: that node is the only element of set $X$ and all the other vertices of the graph are in $Y$ (Figure 1B). Because of this phenomenon, the minimum cut is frequently queried for the ``balanced'' case, when the size (i.e., the number of vertices) of $X$ and $Y$ needs to be equal (or, more exactly, may differ by at most one if the number of the vertices of the graph is odd), see Figure 1C. This problem is referred to as {\em the balanced minimum cut} or the {\em minimum bisection} problem. If the minimum bisection is small that means that there exist a partition of the vertices into two sets of equal size that are connected with only a few edges. If the minimum bisection is large then the two half-sets in {\em every possible bisections} of the graph are connected by many edges.

Therefore, the balanced minimum cut of a graph is independent of the particular labeling of the nodes. The number of all the balanced cuts in a graph with $n$ vertices is greater than 
$${1\over n+1}2^n,$$
that is, for $n=250$, this number is very close to the number of atoms in the visible universe \cite{Ade2013}. Consequently, one cannot practically compute the minimum bisecton width by reviewing all the bisectons in a graph of that size. Moreover, the complexity of computing this quantity is known to be NP-hard \cite{Garey1976} in general, but with contemporary integral programming approaches, for the graph-sizes we are dealing with, the exact values are computable.

In computer engineering, an important measure of the quality of an interconnection network is its minimum bisection width \cite{Tarjan1983a}: the higher the width is the better the network.

For the whole brain graph, as it is anticipated, we have found that the minimum balanced cut is almost exactly represents the edges crossing the {\em corpus callosum}, connecting the two cerebral hemispheres. 

We show that within both hemispheres, the minimum bisection size of female connectomes are significantly larger than the minimum bisection size of the males. Much more importantly, we show that this remains true if we {\em normalize with the  sum of all edge-weights}: that is, {\em this phenomenon cannot be due} to the higher number of edges or the greater edge weights in the female brain: it is an intrinsic property of the female brain graph in our data analyzed.

For example, in the 234-vertex resolution, in the left hemisphere, the normalized balanced minimum cut in females, on the average, is $0.09416$, in the males $0.07896$, $p=0.00153$ (see the Appendix for tables with the results).

We think that this finding is one of the main results of the present work: even if the significant difference in the weighted edge numbers are due to some artifacts in the data acquisition/processing workflow, the normalized balanced minimum cut size seems to be independent from those processes.

\begin{figure} [h!]
	\centering
	\includegraphics[width=5.2in]{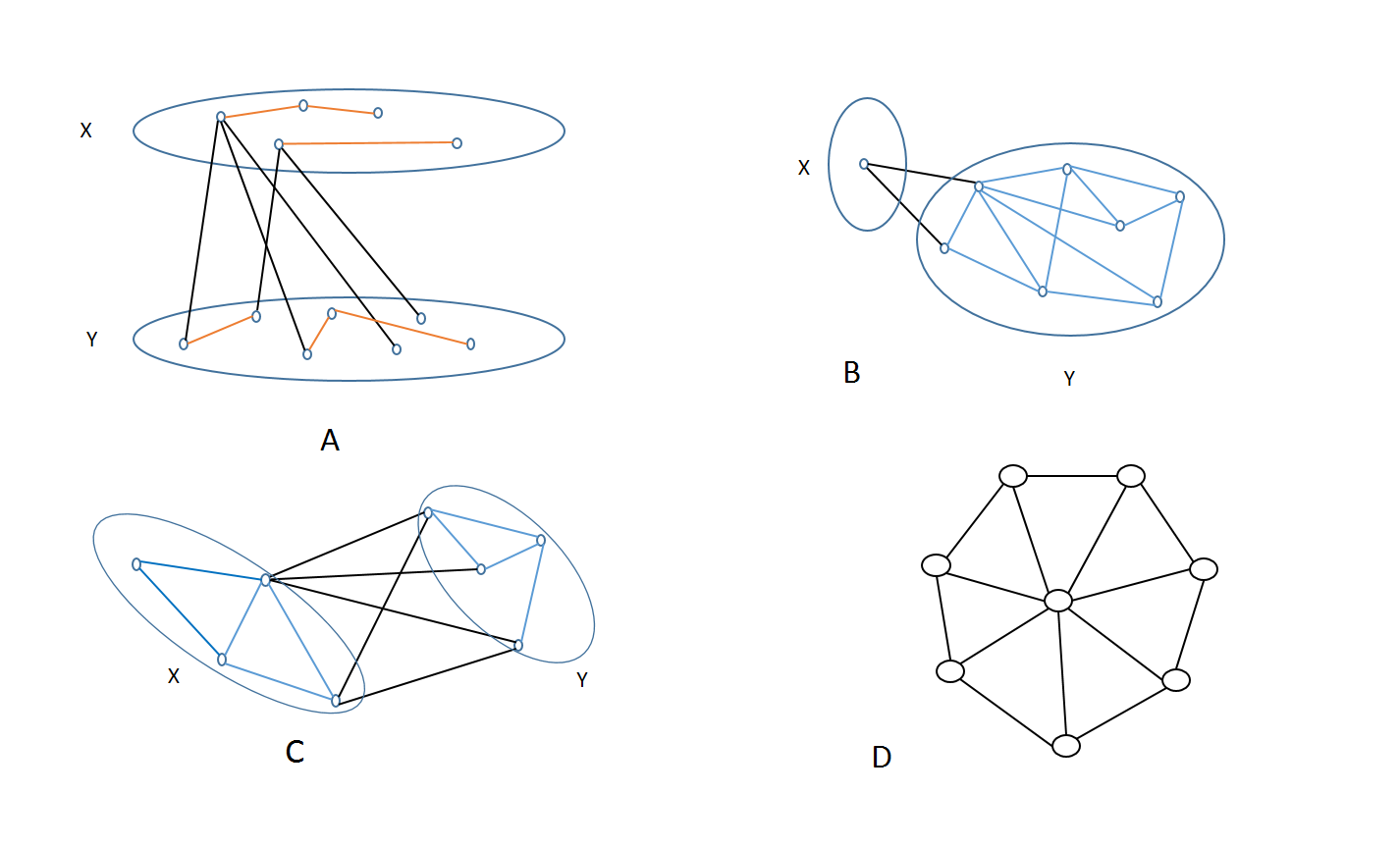}
	\caption{Panel A: An X-Y cut. The cut-edges are colored black. Panel B: An un-balanced minimum cut. Panel C: A balanced cut. Panel D: The wheel graph.  }
\end{figure}

\subsection{Eigengap and the expander property} Expander graphs and the expander-property of graphs are one of the most interesting area of graph theory: they are closely related to the convergence rate and the ergodicity of Markov chains, and have applications in the design of communication- and sorting networks and methods for de-randomizing algorithms  \cite{Hoory2006}.  A graph is an $\varepsilon$-expander, if every -- not too small and not too large -- vertex-set $S$ of the graph has at least $\varepsilon|S|$ outgoing edges (see \cite{Hoory2006} for the exact definition). 

Random walks on good expander graphs converge very fast to the limit distribution: this means that good expander graphs, in a certain sense, are ``intrinsically better'' connected  than bad expanders. It is known that large eigengap of the walk transition matrix of the graph implies good expansion property \cite{Hoory2006}.

We have found that women's connectomes have significantly larger eigengap, and, consequently, they are better expander graphs than the connectomes of men. For example, in the 83-node resolution, in the left hemisphere and in the unweighted graph, the average female connectome's eigengap is  $0.306$ while in the case of men it is $0.272$, with $p=0.00458$.

\subsection{The number of spanning forests}
A {\em tree} in graph theory is a connected, cycle-free graph. Any tree on $n$ vertices has the same number of edges: $n-1$. Trees, and tree-based structures are common in science: phylogenetic trees, hierarchical clusters, data-storage on hard-disks, or a computational model called {\em decision trees} all apply graph-theoretic trees. A {\em spanning tree} is a minimal subgraph of a connected graph that is still connected. Some graphs have no spanning trees at all: only connected graphs have spanning trees. A tree has only one spanning tree: itself. Any connected graph on $n$ vertices has a minimum of $n-1$ and a maximum of $n(n-1)/2$ edges \cite{Lovasz2007}. A connected graph with few edges still may have exponentially many different spanning trees: e.g., the $n$-vertex wheel on Figure 1D has at least $2^{n-1}$ spanning trees (for $n\geq 4$). Cayley's famous theorem, and its celebrated proof with Prüfer codes \cite{Pruefer1918} shows that the number of spanning trees of the complete graph on $n$ vertices is $n^{n-2}$. 

If a graph is not connected, then it contains more than one connected components. Each connected component has at least one spanning tree, and the whole graph has at least one {\em spanning forest}, comprising of the spanning trees of the components. The number of spanning forests is clearly the product of the numbers of the spanning trees of the components.

For graphs in general, one can compute the number of their spanning forests by Kirchoff's matrix tree theorem \cite{Kirchoff1847,Chung1997} using the eigenvalues of the Laplacian matrix \cite{Chung1997} of the graph. 

We show that female connectomes have significantly higher number of spanning trees than the connectomes of males. For example, in the 129-vertex resolution, in the left hemisphere, the logarithm of the number of the spanning forests in the unweighted case are $162.01$ in females, $158.88$ in males with $p=0.013$.

\section{Materials and Methods} 

\subsection{Data source and graph computation:} The dataset applied is a subset of the Human Connectome Project \cite{McNab2013} anonymized 500 Subjects Release:

 \noindent (http://www.humanconnectome.org/documentation/S500) of healthy subjects between 22 and 35 years of age. Data was downloaded in October, 2014. The Connectome Mapper Toolkit \cite{Daducci2012} (http://cmtk.org) was applied for brain tissue segmentation into grey and white matter, partitioning, tractography and the construction of the graphs from the fibers identified in the tractography step. The Connectome Mapper Toolkit \cite{Daducci2012} default partitioning was used (computed by the FreeSurfer, and based on the Desikan-Killiany anatomical atlas) into 83, 129 and 234 cortical and sub-cortical structures (as the brainstem and deep-grey nuclei), referred to as ``Regions of Interest'', ROIs, (see Figure 4 in \cite{Daducci2012}). 
 Tractography was performed by choosing the deterministic streamline method \cite{Daducci2012} with randomized seeding.

 The graphs were constructed as follows: the nodes correspond to the ROIs in the specific resolution. Two nodes were connected by an edge if there exists at least one fiber (determined by the tractography step) connecting the ROIs, corresponding to the nodes. More than one fibers, connecting the same nodes, may give rise to the weight of that edge, depending on the weighting method. Loops were deleted from the graph.

The weights of the edges are assigned by several methods, taking into account the lengths and the multiplicities of the fibers, connecting the nodes:

\begin{itemize}
	\item \texttt{Unweighted}: Each edge has weight 1.
	\item \texttt{FiberN}: The number of fibers traced along the edge: this number is larger than one if more than one fibers connect two cortical or sub-cortical areas, corresponding to the two endpoints of the edge.
	\item \texttt{FAMean}: The arithmetic mean of the fractional anisotropies \cite{Basser2011} of the fibers, belonging to the edge.
	\item \texttt{FiberLengthMean}: The average length of the fibers, connecting the two endpoints of the edge.
	\item \texttt{FiberNDivLength}: The number of fibers belonging to the edge, divided by their average length. This quantity is related to the simple electrical model of the nerve fibers: by modeling the fibers as electrical resistors with resistances proportional to the average fiber length, this quantity is precisely the conductance between the two regions of interest. Additionally, \texttt{FiberNDivLength} can be observed as a reliability measure of the edge: longer fibers are less reliable than the shorter ones, due to possible error accumulation in the tractography algorithm that constructs the fibers from the anisotropy data. Multiple fibers connecting the same two ROIs, corresponding to the endpoints, add to the reliability of the edge, because of the independently tractographed connections.
\end{itemize}

 By {\em generalized adjacency matrix} we mean a matrix of size $n\times n$ where $n$ is the number of {\em nodes} (or {\em vertices}) in the graph, whose rows and columns correspond to the nodes, and whose each element is either zero if there is no edge between the two nodes, or equals to the weight of the edge connecting the two nodes. By the {\em generalized degree} of a node we mean the sum of the weights of the edges adjacent to that node. Note that the generalized degree of the node $v$ is exactly the sum of the elements in the row (or column) of the generalized adjacency matrix corresponding to $v$. By {\em generalized Laplacian matrix} we mean the matrix $D - A$, where $D$ is a diagonal matrix containing the generalized degrees, and $A$ is the generalized adjacency matrix.

\subsection{Graph parameters:}

We calculated various graph parameters for each brain graph and weight function. These parameters included:

\begin{itemize}
	\item Number of edges (\texttt{Sum}). The weighted version of this quantity is the sum of the weights of the edges.
	
	\item Normalized largest eigenvalue (\texttt{AdjLMaxDivD}): The largest eigenvalue of the generalized adjacency matrix, divided by the average degree. Dividing by the average degree of vertices was necessary because the largest eigenvalue is bounded by the average- and maximum degrees, and thus is considered by some a kind of ``average degree'' itself \cite{Lovasz2007}. This means that a denser graph may have a bigger $\lambda_{max}$ largest eigenvalue solely because of a larger average degree. We note that the average degree is already defined by the sum of weights.

	\item Eigengap of the transition matrix (\texttt{PGEigengap}): The transition matrix $P_G$ is obtained by dividing all the rows of the generalized adjacency matrix by the generalized degree of the corresponding node. When performing a random walk on the graph, for nodes $i$ and $j$, the corresponding matrix element describes the probability of transitioning to node $j$, supposing that we are at node $i$. The eigengap of a matrix is the difference of the largest and the second largest eigenvalue. It is characteristic to the expander properties of the graph: the larger the gap, the better expander is the graph (see \cite{Hoory2006} for the exact statements and proofs). 
	
	\item Hoffman's bound (\texttt{HoffmanBound}): The expression $$1 + \frac{\lambda_{max}}{|\lambda_{min}|},$$ where $\lambda_{max}$ and $\lambda_{min}$ denote the largest and smallest eigenvalues of the adjacency matrix. It is a lower bound for the chromatic number of the graph. The chromatic number is generally higher for denser graphs, as the addition of an edge may make a previously valid coloring invalid.
	
	\item Logarithm of number of spanning forests (\texttt{LogAbsSpanningForestN}): The number of the spanning trees in a connected graph can be calculated from the spectrum of its Laplacian \cite{Kirchoff1847,Chung1997}.  Denser graphs tend to have more spanning trees, as the addition of an edge introduces zero or more new spanning trees. If a graph is not connected, then the number of spanning forests is the product of the numbers of the spanning trees of the components. The parameter \texttt{LogAbsSpanningForestN} equals to the logarithm of the number of spanning forests in the unweighted case. In the case of other weight functions, if we define the weight of a tree by the product of the weights of its edges, then this parameter equals to the sum of the logarithms of the weights of the spanning trees in the forests.	
	
	\item Balanced minimum cut, divided by the number of edges (\texttt{MinCutBalDivSum}): The task is to partition the graph into two sets whose size may differ from each other by at most 1, so that the number of edges crossing the cut is minimal. This is the ``balanced minimum cut'' problem, or sometimes called the ``minimum bisection width'' problem. For the whole brain graph, our expectation was that the minimum cut corresponds to the boundary of the two hemispheres, which was indeed proven when we analyzed the results.
	
	\item Minimum cost spanning tree (\texttt{MinSpanningForest}), calculated with Kruskal's algorithm. 
	
	\item Minimum weighted vertex cover (\texttt{MinVertexCover}): Each vertex should have a (possibly fractional) weight assigned such that, for each edge, the sum of the weights of its two endpoints is at least 1. This is the fractional relaxation of the NP-hard vertex-cover problem \cite{Hochbaum1982}. The minimum of the sum of all vertex-weights is computable by a linear programming approach.
	
	\item Minimum vertex cover (\texttt{MinVertexCoverBinary}): Same as above, but each weight must be 0 or 1. In other words, a minimum size set of vertices is selected such that each edge is covered by at least one of the selected vertices. This NP-hard graph-parameter is computed only for the unweighted case. The exact values are computed by an integer programming solver SCIP (http://scip.zib.de), \cite{Achterberg2008, Achterberg2009}.

\end{itemize}

The above 9 parameters were computed for all three resolutions and for the left and the right hemispheres and also for the whole connectome, with all 5 weight functions (with the following exceptions: \texttt{MinVertexCoverBinary} was computed only for the unweighted case, and the \texttt{MinSpanningForest} was not computed for the unweighted case).

\subsection{Statistical analysis}

Since each connectome was computed in multiple resolutions (in 83, 129 and 234 nodes), we had three graphs for each brain. In addition, the parameters were calculated separately for the connectome within the left and right hemispheres as well, not only the whole graph, since we intended to examine whether statistically significant differences can be attributed to the left or right hemispheres. Each subjects' brain was corresponded to 9 graphs (3 resolutions, each in the left and the right hemispheres, plus the whole cortex with sub-cortical areas) and for each graph we calculated 9 parameters, each (with the exceptions noted above) with 5 different edge weights. This means that we assigned $7\cdot5\cdot3+1\cdot3+4\cdot3=120$ attributes to each resolution of the 96 brains, that is, 360 attributes to each brain.

The statistical null hypothesis \cite{Hoel1984} of ours was that the graph parameters do not differ between the male and the female groups. As the first approach, we have used ANOVA (Analysis of variance) \cite{Wonnacott1972} to assign p-values for all parameters in each hemispheres and in each resolutions and in each weight-assignments. 

Our very large number of attributes may lead to false negatives, i.e., to ``type II'' statistical errors: in other words, it may happen that an attribute, with a very small p-value may appear ``at random'', simply because we tested a lot of attributes. In order to deal with ``type II'' statistical errors, we followed the route described below.

We divided the population randomly into two sets by the parity of the sum of the digits in their ID. The first set was used for making hypotheses and the second set for testing these hypotheses. This was necessary to avoid type II errors resulting from multiple testing correction. If we made hypotheses for all the numerical parameters, then the Holm-Bonferroni correction \cite{Holm1979} we used would have unnecessarily increased the p-values. Thus we needed to filter the hypotheses first, and that is why we needed the first set. Testing on the first set allowed us to reduce the number of hypotheses and test only a few of them on the second set.

The hypotheses were filtered by performing ANOVA (Analysis of variance) \cite{Wonnacott1972} on the first set. Only those hypotheses were selected to qualify for the second round where the p-value was less than 1\%.  The selected hypotheses were then tested for the second set as well, and the resulting p-value corrected with the Holm-Bonferroni correction method \cite{Holm1979} with a significance level of 5\%. 

In Table 1 those hypotheses rejected were highlighted in bold, meaning that {\em all} the corresponding graph parameters differ significantly in sex groups at a combined significance level of 5\%. 

We also highlighted (in italic) those p-values which were individually less than the threshold, meaning that these hypotheses can {\em individually} be rejected at a level of 5\%, but it is very likely that {\em not all} of these graph parameters are significantly different between the sexes.

\section{Conclusions:} We have computed 83-, 129- and 234-vertex-graphs from the diffusion MRI images of the 96 subjects of 52 females and 44 males, between the age of 22 and 35. We have found, after a careful statistical analysis, significant differences between some graph theoretical parameters of the male and female brain graphs. Our findings show that the female brain graphs have generally more edges (counted with and without weights), have larger normalized minimum bisection widths and have more spanning trees (counted with and without weights) than the connectomes of males (Table 1). Additionally, with weaker statistical validity, some spectral properties and the minimum vertex cover also differ in the connectomes of different sexes (each with $p<0.02$).

\section{Data availability:} The unprocessed and pre-processed MRI data is available at the Human Connectome Project's website:

 http://www.humanconnectome.org/documentation/S500 \cite{McNab2013}. 

\subsection{Table 1}
{\small
\begin{longtable}{l | cccc}
	Scale & Property & p (1st) & p (2nd) & p (corrected) \\ 
	129 & Right\_MinCutBalDivSum\_FAMean & 0.00807 & {\em 0.00003} & \textbf{0.00401} \\ 
	83 & All\_LogSpanningForestN\_FiberNDivLength & 0.00003 & {\em 0.00004} & \textbf{0.00451} \\ 
	234 & All\_PGEigengap\_FiberNDivLength & 0.00321 & {\em 0.00007} & \textbf{0.00798} \\ 
	129 & All\_PGEigengap\_FiberNDivLength & 0.00792 & {\em 0.00011} & \textbf{0.01303} \\ 
	83 & Left\_MinCutBalDivSum\_FiberN & 0.00403 & {\em 0.00011} & \textbf{0.01300} \\ 
	83 & Right\_MinCutBalDivSum\_FAMean & 0.00496 & {\em 0.00015} & \textbf{0.01744} \\ 
	129 & Left\_PGEigengap\_FiberNDivLength & 0.00223 & {\em 0.00015} & \textbf{0.01797} \\ 
	234 & All\_PGEigengap\_FiberN & 0.00826 & {\em 0.00022} & \textbf{0.02517} \\ 
	83 & All\_Sum\_Unweighted & 0.00025 & {\em 0.00022} & \textbf{0.02504} \\ 
	129 & Left\_MinCutBalDivSum\_FiberN & 0.00001 & {\em 0.00023} & \textbf{0.02563} \\ 
	83 & All\_LogSpanningForestN\_FiberN & 0.00001 & {\em 0.00028} & \textbf{0.03084} \\ 
	83 & Right\_Sum\_FAMean & 0.00028 & {\em 0.00029} & \textbf{0.03224} \\ 
	234 & All\_Sum\_Unweighted & 0.00063 & {\em 0.00032} & \textbf{0.03512} \\ 
	234 & Left\_PGEigengap\_FiberNDivLength & 0.00013 & {\em 0.00038} & \textbf{0.04171} \\ 
	129 & All\_Sum\_Unweighted & 0.00026 & {\em 0.00042} & \textbf{0.04563} \\ 
	234 & All\_Sum\_FAMean & 0.00014 & {\em 0.00047} & \textbf{0.04988} \\ 
	129 & All\_LogSpanningForestN\_FiberN & 0.00000 & {\em 0.00048} & 0.05045 \\ 
	83 & All\_Sum\_FAMean & 0.00029 & {\em 0.00050} & 0.05260 \\ 
	129 & Right\_Sum\_FAMean & 0.00062 & {\em 0.00051} & 0.05355 \\ 
	234 & Right\_PGEigengap\_FiberNDivLength & 0.00041 & {\em 0.00053} & 0.05414 \\ 
	83 & Left\_Sum\_Unweighted & 0.00378 & {\em 0.00068} & 0.06936 \\ 
	234 & Right\_Sum\_FAMean & 0.00085 & {\em 0.00084} & 0.08454 \\ 
	234 & Left\_Sum\_Unweighted & 0.00293 & {\em 0.00092} & 0.09212 \\ 
	129 & All\_Sum\_FAMean & 0.00015 & {\em 0.00097} & 0.09650 \\ 
	234 & Left\_MinCutBalDivSum\_FiberN & 0.00002 & {\em 0.00108} & 0.10539 \\ 
	83 & Left\_LogSpanningForestN\_FiberNDivLength & 0.00343 & {\em 0.00116} & 0.11274 \\ 
	83 & All\_LogSpanningForestN\_Unweighted & 0.00113 & {\em 0.00121} & 0.11629 \\ 
	234 & Left\_MinCutBalDivSum\_FiberLengthMean & 0.00411 & {\em 0.00123} & 0.11646 \\ 
	83 & All\_LogSpanningForestN\_FAMean & 0.00012 & {\em 0.00126} & 0.11823 \\ 
	83 & Right\_Sum\_Unweighted & 0.00019 & {\em 0.00128} & 0.11891 \\ 
	129 & Left\_MinCutBalDivSum\_Unweighted & 0.00265 & {\em 0.00134} & 0.12351 \\ 
	83 & Left\_MinCutBalDivSum\_Unweighted & 0.00206 & {\em 0.00136} & 0.12370 \\ 
	129 & Left\_PGEigengap\_FiberN & 0.00382 & {\em 0.00142} & 0.12775 \\ 
	234 & All\_LogSpanningForestN\_FAMean & 0.00043 & {\em 0.00150} & 0.13343 \\ 
	234 & Left\_PGEigengap\_FiberN & 0.00066 & {\em 0.00163} & 0.14369 \\ 
	129 & Right\_LogSpanningForestN\_FAMean & 0.00143 & {\em 0.00170} & 0.14769 \\ 
	83 & Left\_MinCutBalDivSum\_FiberNDivLength & 0.00031 & {\em 0.00175} & 0.15023 \\ 
	129 & All\_LogSpanningForestN\_FiberNDivLength & 0.00000 & {\em 0.00177} & 0.15009 \\ 
	129 & All\_LogSpanningForestN\_Unweighted & 0.00218 & {\em 0.00182} & 0.15279 \\ 
	129 & Right\_Sum\_Unweighted & 0.00068 & {\em 0.00186} & 0.15417 \\ 
	129 & Left\_PGEigengap\_FAMean & 0.00995 & {\em 0.00191} & 0.15694 \\ 
	129 & All\_LogSpanningForestN\_FAMean & 0.00019 & {\em 0.00211} & 0.17093 \\ 
	234 & Left\_Sum\_FAMean & 0.00026 & {\em 0.00212} & 0.16978 \\ 
	83 & Right\_LogSpanningForestN\_FAMean & 0.00067 & {\em 0.00239} & 0.18842 \\ 
	234 & Left\_PGEigengap\_FAMean & 0.00141 & {\em 0.00240} & 0.18684 \\ 
	83 & Left\_PGEigengap\_Unweighted & 0.00458 & {\em 0.00243} & 0.18738 \\ 
	129 & Left\_MinCutBalDivSum\_FiberLengthMean & 0.00892 & {\em 0.00245} & 0.18596 \\ 
	83 & Left\_Sum\_FAMean & 0.00056 & {\em 0.00279} & 0.20893 \\ 
	234 & Left\_MinCutBalDivSum\_Unweighted & 0.00154 & {\em 0.00289} & 0.21355 \\ 
	234 & Left\_PGEigengap\_FiberLengthMean & 0.00554 & {\em 0.00295} & 0.21516 \\ 
	234 & Right\_LogSpanningForestN\_FAMean & 0.00380 & {\em 0.00305} & 0.21935 \\ 
	234 & Left\_PGEigengap\_Unweighted & 0.00176 & {\em 0.00338} & 0.24029 \\ 
	83 & Left\_PGEigengap\_FAMean & 0.00215 & {\em 0.00359} & 0.25152 \\ 
	83 & Left\_LogSpanningForestN\_FiberN & 0.00012 & {\em 0.00395} & 0.27269 \\ 
	129 & Left\_Sum\_Unweighted & 0.00232 & {\em 0.00456} & 0.31006 \\ 
	83 & Left\_LogSpanningForestN\_FAMean & 0.00082 & {\em 0.00496} & 0.33212 \\ 
	234 & Right\_MinCutBalDivSum\_Unweighted & 0.00462 & {\em 0.00543} & 0.35825 \\ 
	83 & Right\_LogSpanningForestN\_FiberNDivLength & 0.00022 & {\em 0.00587} & 0.38180 \\ 
	234 & Left\_LogSpanningForestN\_FAMean & 0.000129 & {\em 0.00595} & 0.38054 \\ 
	234 & Right\_PGEigengap\_Unweighted & 0.00095 & {\em 0.00626} & 0.39459 \\ 
	129 & Left\_Sum\_FAMean & 0.00032 & {\em 0.00660} & 0.40907 \\ 
	83 & Left\_AdjLMaxDivD\_FiberN & 0.00501 & {\em 0.00804} & 0.49040 \\ 
	234 & Right\_Sum\_Unweighted & 0.00224 & {\em 0.00845} & 0.50692 \\ 
	234 & Right\_PGEigengap\_FiberN & 0.00009 & {\em 0.00910} & 0.53671 \\ 
	129 & All\_Sum\_FiberN & 0.00000 & {\em 0.00938} & 0.54418 \\ 
	234 & Right\_PGEigengap\_FAMean & 0.00074 & {\em 0.00974} & 0.55538 \\ 
	129 & Right\_PGEigengap\_FAMean & 0.00296 & {\em 0.00981} & 0.54933 \\ 
	83 & Right\_PGEigengap\_Unweighted & 0.00087 & {\em 0.01053} & 0.57889 \\ 
	129 & Right\_MinCutBalDivSum\_FiberN & 0.00563 & {\em 0.01101} & 0.59432 \\ 
	129 & Right\_MinCutBalDivSum\_Unweighted & 0.00492 & {\em 0.01212} & 0.64227 \\ 
	129 & Left\_LogSpanningForestN\_FAMean & 0.00106 & {\em 0.01218} & 0.63359 \\ 
	129 & Left\_LogSpanningForestN\_FiberN & 0.00014 & {\em 0.01258} & 0.64134 \\ 
	83 & All\_Sum\_FiberN & 0.00000 & {\em 0.01290} & 0.64480 \\ 
	234 & All\_Sum\_FiberN & 0.00000 & {\em 0.01358} & 0.66520 \\ 
	83 & Right\_LogSpanningForestN\_Unweighted & 0.00541 & {\em 0.01438} & 0.69010 \\ 
	129 & Left\_LogSpanningForestN\_FiberNDivLength & 0.00288 & {\em 0.01447} & 0.67995 \\ 
	129 & Right\_PGEigengap\_Unweighted & 0.00242 & {\em 0.01676} & 0.77084 \\ 
	129 & Right\_PGEigengap\_FiberN & 0.00869 & {\em 0.01706} & 0.76750 \\ 
	234 & All\_MinVertexCover\_FAMean & 0.00289 & {\em 0.01713} & 0.75373 \\ 
	83 & All\_HoffmanBound\_FAMean & 0.00087 & {\em 0.02011} & 0.86462 \\ 
	83 & All\_Sum\_FiberNDivLength & 0.00002 & {\em 0.02117} & 0.88929 \\ 
	234 & Right\_MinCutBalDivSum\_FiberN & 0.00234 & {\em 0.02197} & 0.90065 \\ 
	83 & Right\_LogSpanningForestN\_FiberN & 0.00083 & {\em 0.02539} & 1.01567 \\ 
	234 & Right\_MinCutBalDivSum\_FiberLengthMean & 0.00234 & {\em 0.02663} & 1.03841 \\ 
	83 & Right\_MinCutBalDivSum\_FiberNDivLength & 0.00072 & {\em 0.02854} & 1.08446 \\ 
	129 & Left\_MinCutBalDivSum\_FiberNDivLength & 0.00019 & {\em 0.02897} & 1.07195 \\ 
	83 & Right\_PGEigengap\_FAMean & 0.00112 & {\em 0.02948} & 1.06119 \\ 
	234 & All\_LogSpanningForestN\_FiberN & 0.00091 & {\em 0.03308} & 1.15795 \\ 
	234 & Right\_PGEigengap\_FiberLengthMean & 0.00367 & {\em 0.03369} & 1.14542 \\ 
	129 & Right\_MinCutBalDivSum\_FiberLengthMean & 0.00768 & {\em 0.04500} & 1.48511 \\ 
	129 & All\_Sum\_FiberNDivLength & 0.00008 & {\em 0.04728} & 1.51293 \\ 
	129 & Right\_LogSpanningForestN\_FiberNDivLength & 0.00051 & {\em 0.04891} & 1.51627 \\ 
	234 & All\_LogSpanningForestN\_FiberNDivLength & 0.00106 & 0.05095 & 1.52842 \\ 
	129 & Right\_LogSpanningForestN\_FiberN & 0.00045 & 0.05578 & 1.61751 \\ 
	83 & Right\_MinCutBalDivSum\_FiberN & 0.00346 & 0.06284 & 1.75951 \\ 
	83 & Right\_HoffmanBound\_FiberNDivLength & 0.005129 & 0.06309 & 1.70341 \\ 
	83 & Right\_PGEigengap\_FiberLengthMean & 0.00949 & 0.06515 & 1.69395 \\ 
	234 & Left\_MinCutBalDivSum\_FiberNDivLength & 0.00642 & 0.06548 & 1.63696 \\ 
	234 & Left\_MinVertexCover\_FAMean & 0.00107 & 0.07139 & 1.71336 \\ 
	234 & All\_Sum\_FiberNDivLength & 0.00044 & 0.07318 & 1.68305 \\ 
	83 & Right\_Sum\_FiberN & 0.00000 & 0.07799 & 1.71586 \\ 
	83 & Right\_Sum\_FiberNDivLength & 0.00018 & 0.07920 & 1.66329 \\ 
	129 & Left\_Sum\_FiberN & 0.00000 & 0.08380 & 1.67598 \\ 
	129 & Right\_Sum\_FiberN & 0.00001 & 0.08653 & 1.64406 \\ 
	129 & Left\_HoffmanBound\_Unweighted & 0.00848 & 0.08944 & 1.60984 \\ 
	83 & Left\_Sum\_FiberN & 0.00000 & 0.09430 & 1.60310 \\ 
	234 & Left\_Sum\_FiberN & 0.00040 & 0.11447 & 1.83157 \\ 
	129 & Right\_Sum\_FiberNDivLength & 0.00180 & 0.12102 & 1.81523 \\ 
	234 & Right\_Sum\_FiberN & 0.00012 & 0.16411 & 2.29752 \\ 
	83 & Left\_Sum\_FiberNDivLength & 0.00043 & 0.16774 & 2.18062 \\ 
	129 & Left\_Sum\_FiberNDivLength & 0.00100 & 0.22542 & 2.70502 \\ 
	234 & Right\_Sum\_FiberNDivLength & 0.00562 & 0.23691 & 2.60604 \\ 
	83 & Right\_HoffmanBound\_FAMean & 0.00587 & 0.32069 & 3.20692 \\ 
	83 & All\_MinVertexCoverBinary\_Unweighted & 0.00716 & 0.38829 & 3.49459 \\ 
	234 & Right\_LogSpanningForestN\_FiberNDivLength & 0.00940 & 0.40996 & 3.27971 \\ 
	83 & Left\_HoffmanBound\_FiberN & 0.00175 & 0.41913 & 2.93394 \\ 
	83 & All\_MinVertexCover\_FiberNDivLength & 0.00036 & 0.46677 & 2.80065 \\ 
	83 & Right\_MinSpanningForest\_FiberLengthMean & 0.00491 & 0.55239 & 2.76195 \\ 
	234 & Right\_MinSpanningForest\_FiberLengthMean & 0.00601 & 0.55631 & 2.22523 \\ 
	129 & All\_MinVertexCover\_FiberN & 0.00232 & 0.71406 & 2.14217 \\ 
	83 & All\_MinVertexCover\_FiberN & 0.00244 & 0.84437 & 1.68874 \\ 
	234 & All\_MinVertexCover\_FiberN & 0.00055 & 0.92958 & 0.92958 \\ 
	\\
\caption{The results and the statistical analysis of the graph-theoretical evaluation of the sex differences in the 96 diffusion MRI images. The first column gives the resolution in each hemisphere; the number of nodes in the whole graph is 83, 129 and 234, respectively. The second column describes the graph parameter computed: its syntactics is as follows: each parameter-name contains two separating ``$\_$'' symbols that define three parts of the parameter-name. The first part describe the hemisphere or the whole connectome with the words Left, Right or All. The second part describes the parameter computed, and the third part the weight function used (their definitions are given in section ``Materials and methods''). The third column contains the p-values of the first round, the second column the p-values of the second round, and the third column the (very strict) Holm-Bonferroni correction of the p-value. With p=0.05 {\em all} the first 12 rows describe significantly different graph theoretical properties between sexes. One-by-one, each row with italic third column describe significant differences between sexes, with p=0.05. For the details we refer to the section ``Statistical analysis''. }
\end{longtable}
}

\section{Acknowledgments}
The authors declare no conflicts of interest.



\vfil
\eject

\section*{Appendix}

In this appendix we list the graph-theoretic parameters computed for the resolutions of  83, 129 and 234 vertex graphs. The tables contain their arithmetic means in the male and female groups, and the corresponding p-values. The values in these tables contain the values corresponded to round 1 (see the ``Statistical analysis'' subsection in the main text). 

The graph-parameters are defined in the caption of Table 1.

Significant differences ($p<0.01$) are denoted with an asterisk in the last column.

\scriptsize{
	\subsection*{Scale 83, round 1}
		\begin{longtable}{l | cccc}	
	Property & Female & Male & p-value &  \\ 
	All\_AdjLMaxDivD\_FAMean & 1.36008 & 1.37750 & 0.06806 \\ 
	All\_AdjLMaxDivD\_FiberLengthMean & 1.44214 & 1.43602 & 0.72030 \\ 
	All\_AdjLMaxDivD\_FiberN & 2.02416 & 2.10529 & 0.05606 \\ 
	All\_AdjLMaxDivD\_FiberNDivLength & 1.84476 & 1.86864 & 0.41834 \\ 
	All\_AdjLMaxDivD\_Unweighted & 1.26760 & 1.26456 & 0.63251 \\ 
	All\_HoffmanBound\_FAMean & 4.36096 & 4.18564 & 0.00087 & $*$ \\ 
	All\_HoffmanBound\_FiberLengthMean & 3.21938 & 3.26552 & 0.33136 \\ 
	All\_HoffmanBound\_FiberN & 2.63525 & 2.55573 & 0.03144 \\ 
	All\_HoffmanBound\_FiberNDivLength & 2.51038 & 2.40550 & 0.01815 \\ 
	All\_HoffmanBound\_Unweighted & 4.55192 & 4.43931 & 0.04616 \\ 
	All\_LogSpanningForestN\_FAMean & 110.69890 & 101.82758 & 0.00012 & $*$ \\ 
	All\_LogSpanningForestN\_FiberLengthMean & 456.60084 & 452.95875 & 0.18687 \\ 
	All\_LogSpanningForestN\_FiberN & 397.53780 & 389.79037 & 0.00001 & $*$ \\ 
	All\_LogSpanningForestN\_FiberNDivLength & 148.03174 & 139.85355 & 0.00003 & $*$ \\ 
	All\_LogSpanningForestN\_Unweighted & 191.66035 & 187.85180 & 0.00113 & $*$ \\ 
	All\_MinCutBalDivSum\_FAMean & 0.00793 & 0.00474 & 0.14869 \\ 
	All\_MinCutBalDivSum\_FiberLengthMean & 0.03115 & 0.02889 & 0.47008 \\ 
	All\_MinCutBalDivSum\_FiberN & 0.02924 & 0.02711 & 0.34092 \\ 
	All\_MinCutBalDivSum\_FiberNDivLength & 0.02868 & 0.02644 & 0.38768 \\ 
	All\_MinCutBalDivSum\_Unweighted & 0.04001 & 0.03721 & 0.28887 \\ 
	All\_MinSpanningForest\_FAMean & 19.78188 & 18.63722 & 0.02232 \\ 
	All\_MinSpanningForest\_FiberLengthMean & 1096.37958 & 1112.97289 & 0.10506 \\ 
	All\_MinSpanningForest\_FiberN & 99.53846 & 102.93333 & 0.14280 \\ 
	All\_MinSpanningForest\_FiberNDivLength & 3.65548 & 3.66822 & 0.93669 \\ 
	All\_MinVertexCoverBinary\_Unweighted & 59.80769 & 59.00000 & 0.00716 & $*$ \\ 
	All\_MinVertexCover\_FAMean & 18.73144 & 18.10619 & 0.01699 \\ 
	All\_MinVertexCover\_FiberLengthMean & 2014.06431 & 1955.70824 & 0.37460 \\ 
	All\_MinVertexCover\_FiberN & 2427.21154 & 2315.20000 & 0.00244 & $*$ \\ 
	All\_MinVertexCover\_FiberNDivLength & 110.25657 & 103.59777 & 0.00036 & $*$ \\ 
	All\_MinVertexCover\_Unweighted & 40.90385 & 41.00000 & 0.32897 \\ 
	All\_PGEigengap\_FAMean & 0.05403 & 0.05071 & 0.28914 \\ 
	All\_PGEigengap\_FiberLengthMean & 0.04167 & 0.03891 & 0.43309 \\ 
	All\_PGEigengap\_FiberN & 0.03156 & 0.02829 & 0.03885 \\ 
	All\_PGEigengap\_FiberNDivLength & 0.03470 & 0.03062 & 0.01847 \\ 
	All\_PGEigengap\_Unweighted & 0.05214 & 0.04740 & 0.09708 \\ 
	All\_Sum\_FAMean & 222.01291 & 201.02562 & 0.00029 & $*$ \\ 
	All\_Sum\_FiberLengthMean & 16845.33062 & 15792.24352 & 0.06219 \\ 
	All\_Sum\_FiberN & 11261.65385 & 10237.13333 & 0.00000 & $*$ \\ 
	All\_Sum\_FiberNDivLength & 476.56342 & 433.37987 & 0.00002 & $*$ \\ 
	All\_Sum\_Unweighted & 567.07692 & 539.80000 & 0.00025 & $*$ \\ 
	Left\_AdjLMaxDivD\_FAMean & 1.33644 & 1.35216 & 0.15767 \\ 
	Left\_AdjLMaxDivD\_FiberLengthMean & 1.40515 & 1.38890 & 0.32795 \\ 
	Left\_AdjLMaxDivD\_FiberN & 1.90607 & 2.02087 & 0.00501 & $*$ \\ 
	Left\_AdjLMaxDivD\_FiberNDivLength & 1.71498 & 1.77482 & 0.07539 \\ 
	Left\_AdjLMaxDivD\_Unweighted & 1.24027 & 1.23523 & 0.43598 \\ 
	Left\_HoffmanBound\_FAMean & 4.55406 & 4.38621 & 0.01297 \\ 
	Left\_HoffmanBound\_FiberLengthMean & 3.25098 & 3.28435 & 0.51250 \\ 
	Left\_HoffmanBound\_FiberN & 2.71430 & 2.61098 & 0.00175 & $*$ \\ 
	Left\_HoffmanBound\_FiberNDivLength & 2.66652 & 2.59451 & 0.13782 \\ 
	Left\_HoffmanBound\_Unweighted & 4.73205 & 4.57434 & 0.01379 \\ 
	Left\_LogSpanningForestN\_FAMean & 53.30579 & 48.82905 & 0.00082 & $*$ \\ 
	Left\_LogSpanningForestN\_FiberLengthMean & 229.63370 & 227.32675 & 0.18765 \\ 
	Left\_LogSpanningForestN\_FiberN & 199.27958 & 195.25428 & 0.00012 & $*$ \\ 
	Left\_LogSpanningForestN\_FiberNDivLength & 73.53683 & 69.82889 & 0.00343 & $*$ \\ 
	Left\_LogSpanningForestN\_Unweighted & 95.46307 & 93.39767 & 0.01389 \\ 
	Left\_MinCutBalDivSum\_FAMean & 0.00687 & 0.00320 & 0.17151 \\ 
	Left\_MinCutBalDivSum\_FiberLengthMean & 0.23438 & 0.21147 & 0.01779 \\ 
	Left\_MinCutBalDivSum\_FiberN & 0.13337 & 0.12011 & 0.00403 & $*$ \\ 
	Left\_MinCutBalDivSum\_FiberNDivLength & 0.11057 & 0.09321 & 0.00031 & $*$ \\ 
	Left\_MinCutBalDivSum\_Unweighted & 0.24513 & 0.22019 & 0.00206 & $*$ \\ 
	Left\_MinSpanningForest\_FAMean & 9.57924 & 9.06313 & 0.04242 \\ 
	Left\_MinSpanningForest\_FiberLengthMean & 561.47024 & 560.36391 & 0.87722 \\ 
	Left\_MinSpanningForest\_FiberN & 51.23077 & 53.73333 & 0.26795 \\ 
	Left\_MinSpanningForest\_FiberNDivLength & 1.82447 & 1.89521 & 0.62729 \\ 
	Left\_MinVertexCoverBinary\_Unweighted & 30.23077 & 29.73333 & 0.09601 \\ 
	Left\_MinVertexCover\_FAMean & 9.23616 & 8.88642 & 0.01371 \\ 
	Left\_MinVertexCover\_FiberLengthMean & 1064.27185 & 1027.73430 & 0.35926 \\ 
	Left\_MinVertexCover\_FiberN & 1158.21154 & 1143.46667 & 0.55321 \\ 
	Left\_MinVertexCover\_FiberNDivLength & 54.26322 & 51.17634 & 0.02122 \\ 
	Left\_MinVertexCover\_Unweighted & 20.80769 & 20.83333 & 0.75017 \\ 
	Left\_PGEigengap\_FAMean & 0.33446 & 0.29469 & 0.00215 & $*$ \\ 
	Left\_PGEigengap\_FiberLengthMean & 0.33383 & 0.29287 & 0.01329 \\ 
	Left\_PGEigengap\_FiberN & 0.16980 & 0.15238 & 0.01654 \\ 
	Left\_PGEigengap\_FiberNDivLength & 0.14486 & 0.13413 & 0.02837 \\ 
	Left\_PGEigengap\_Unweighted & 0.30646 & 0.27160 & 0.00458 & $*$ \\ 
	Left\_Sum\_FAMean & 106.64056 & 96.80731 & 0.00056 & $*$ \\ 
	Left\_Sum\_FiberLengthMean & 8629.73791 & 8122.82646 & 0.13250 \\ 
	Left\_Sum\_FiberN & 5514.61538 & 5049.73333 & 0.00000 & $*$ \\ 
	Left\_Sum\_FiberNDivLength & 233.06402 & 213.49323 & 0.00043 & $*$ \\ 
	Left\_Sum\_Unweighted & 282.50000 & 269.06667 & 0.00378 & $*$ \\ 
	Right\_AdjLMaxDivD\_FAMean & 1.32878 & 1.34242 & 0.14511 \\ 
	Right\_AdjLMaxDivD\_FiberLengthMean & 1.39672 & 1.38478 & 0.30191 \\ 
	Right\_AdjLMaxDivD\_FiberN & 2.00803 & 2.09048 & 0.05380 \\ 
	Right\_AdjLMaxDivD\_FiberNDivLength & 1.76990 & 1.81343 & 0.09784 \\ 
	Right\_AdjLMaxDivD\_Unweighted & 1.25268 & 1.24720 & 0.29540 \\ 
	Right\_HoffmanBound\_FAMean & 4.47438 & 4.28666 & 0.00587 & $*$ \\ 
	Right\_HoffmanBound\_FiberLengthMean & 3.33823 & 3.39478 & 0.29902 \\ 
	Right\_HoffmanBound\_FiberN & 2.67311 & 2.57701 & 0.05411 \\ 
	Right\_HoffmanBound\_FiberNDivLength & 2.62635 & 2.48983 & 0.00560 & $*$ \\ 
	Right\_HoffmanBound\_Unweighted & 4.61480 & 4.50726 & 0.03806 \\ 
	Right\_LogSpanningForestN\_FAMean & 52.25642 & 48.14346 & 0.00067 & $*$ \\ 
	Right\_LogSpanningForestN\_FiberLengthMean & 218.25106 & 216.24411 & 0.16431 \\ 
	Right\_LogSpanningForestN\_FiberN & 190.62427 & 187.02757 & 0.00083 & $*$ \\ 
	Right\_LogSpanningForestN\_FiberNDivLength & 69.84080 & 66.17446 & 0.00022 & $*$ \\ 
	Right\_LogSpanningForestN\_Unweighted & 90.24090 & 88.51678 & 0.00541 & $*$ \\ 
	Right\_MinCutBalDivSum\_FAMean & 0.02476 & 0.00851 & 0.00496 & $*$ \\ 
	Right\_MinCutBalDivSum\_FiberLengthMean & 0.24577 & 0.22309 & 0.02216 \\ 
	Right\_MinCutBalDivSum\_FiberN & 0.13346 & 0.12050 & 0.00346 & $*$ \\ 
	Right\_MinCutBalDivSum\_FiberNDivLength & 0.10831 & 0.09357 & 0.00072 & $*$ \\ 
	Right\_MinCutBalDivSum\_Unweighted & 0.23713 & 0.22022 & 0.01629 \\ 
	Right\_MinSpanningForest\_FAMean & 10.30911 & 9.79708 & 0.10419 \\ 
	Right\_MinSpanningForest\_FiberLengthMean & 532.13580 & 547.85331 & 0.00491 & $*$ \\ 
	Right\_MinSpanningForest\_FiberN & 50.76923 & 52.53333 & 0.26282 \\ 
	Right\_MinSpanningForest\_FiberNDivLength & 1.94340 & 1.89232 & 0.58863 \\ 
	Right\_MinVertexCoverBinary\_Unweighted & 29.07692 & 28.73333 & 0.15457 \\ 
	Right\_MinVertexCover\_FAMean & 9.26572 & 9.03965 & 0.12382 \\ 
	Right\_MinVertexCover\_FiberLengthMean & 934.26071 & 897.95882 & 0.23661 \\ 
	Right\_MinVertexCover\_FiberN & 1169.63462 & 1122.93333 & 0.07986 \\ 
	Right\_MinVertexCover\_FiberNDivLength & 53.57144 & 51.50298 & 0.10452 \\ 
	Right\_MinVertexCover\_Unweighted & 20.11538 & 20.26667 & 0.10527 \\ 
	Right\_PGEigengap\_FAMean & 0.32454 & 0.28808 & 0.00112 & $*$ \\ 
	Right\_PGEigengap\_FiberLengthMean & 0.34029 & 0.29461 & 0.00949 & $*$ \\ 
	Right\_PGEigengap\_FiberN & 0.17666 & 0.15912 & 0.02617 \\ 
	Right\_PGEigengap\_FiberNDivLength & 0.15245 & 0.14034 & 0.01613 \\ 
	Right\_PGEigengap\_Unweighted & 0.29582 & 0.26081 & 0.00087 & $*$ \\ 
	Right\_Sum\_FAMean & 105.62164 & 95.26436 & 0.00028 & $*$ \\ 
	Right\_Sum\_FiberLengthMean & 7644.90330 & 7086.91000 & 0.02974 \\ 
	Right\_Sum\_FiberN & 5378.03846 & 4884.66667 & 0.00000 & $*$ \\ 
	Right\_Sum\_FiberNDivLength & 225.94776 & 206.97587 & 0.00018 & $*$ \\ 
	Right\_Sum\_Unweighted & 261.30769 & 248.26667 & 0.00019 & $*$ \\ 
	\end{longtable}
	
	\subsection*{Scale 129, round 1}
	\begin{longtable}{l | cccc}
			Property & Female & Male & p-value &  \\
		All\_AdjLMaxDivD\_FAMean & 1.40519 & 1.42604 & 0.10040 \\ 
		All\_AdjLMaxDivD\_FiberLengthMean & 1.50483 & 1.50158 & 0.87806 \\ 
		All\_AdjLMaxDivD\_FiberN & 2.14552 & 2.22254 & 0.15242 \\ 
		All\_AdjLMaxDivD\_FiberNDivLength & 2.09783 & 2.04782 & 0.32031 \\ 
		All\_AdjLMaxDivD\_Unweighted & 1.30028 & 1.29097 & 0.27278 \\ 
		All\_HoffmanBound\_FAMean & 4.40157 & 4.29660 & 0.02644 \\ 
		All\_HoffmanBound\_FiberLengthMean & 3.19684 & 3.24689 & 0.32568 \\ 
		All\_HoffmanBound\_FiberN & 2.50604 & 2.48884 & 0.64956 \\ 
		All\_HoffmanBound\_FiberNDivLength & 2.34647 & 2.41938 & 0.07720 \\ 
		All\_HoffmanBound\_Unweighted & 4.62935 & 4.51267 & 0.01233 \\ 
		All\_LogSpanningForestN\_FAMean & 194.37749 & 181.03525 & 0.00019 & $*$ \\ 
		All\_LogSpanningForestN\_FiberLengthMean & 739.78985 & 732.55388 & 0.09867 \\ 
		All\_LogSpanningForestN\_FiberN & 599.76631 & 588.61699 & 0.00000 & $*$ \\ 
		All\_LogSpanningForestN\_FiberNDivLength & 210.52236 & 200.75240 & 0.00000 & $*$ \\ 
		All\_LogSpanningForestN\_Unweighted & 322.09324 & 316.62672 & 0.00218 & $*$ \\ 
		All\_MinCutBalDivSum\_FAMean & 0.00668 & 0.00324 & 0.05930 \\ 
		All\_MinCutBalDivSum\_FiberLengthMean & 0.01706 & 0.01607 & 0.56293 \\ 
		All\_MinCutBalDivSum\_FiberN & 0.02658 & 0.02429 & 0.26627 \\ 
		All\_MinCutBalDivSum\_FiberNDivLength & 0.02495 & 0.02258 & 0.30029 \\ 
		All\_MinCutBalDivSum\_Unweighted & 0.02218 & 0.02065 & 0.30082 \\ 
		All\_MinSpanningForest\_FAMean & 30.14746 & 28.58509 & 0.02073 \\ 
		All\_MinSpanningForest\_FiberLengthMean & 1642.68263 & 1664.23693 & 0.07510 \\ 
		All\_MinSpanningForest\_FiberN & 140.23077 & 140.93333 & 0.55077 \\ 
		All\_MinSpanningForest\_FiberNDivLength & 4.42401 & 4.43795 & 0.92181 \\ 
		All\_MinVertexCoverBinary\_Unweighted & 96.46154 & 96.26667 & 0.66793 \\ 
		All\_MinVertexCover\_FAMean & 29.56250 & 28.72424 & 0.02181 \\ 
		All\_MinVertexCover\_FiberLengthMean & 3230.07900 & 3121.21684 & 0.29100 \\ 
		All\_MinVertexCover\_FiberN & 2444.92308 & 2337.40000 & 0.00232 & $*$ \\ 
		All\_MinVertexCover\_FiberNDivLength & 120.18766 & 116.22553 & 0.02502 \\ 
		All\_MinVertexCover\_Unweighted & 63.88462 & 63.96667 & 0.35805 \\ 
		All\_PGEigengap\_FAMean & 0.03143 & 0.02928 & 0.25524 \\ 
		All\_PGEigengap\_FiberLengthMean & 0.02427 & 0.02260 & 0.43054 \\ 
		All\_PGEigengap\_FiberN & 0.02781 & 0.02453 & 0.01902 \\ 
		All\_PGEigengap\_FiberNDivLength & 0.02880 & 0.02498 & 0.00792 & $*$ \\ 
		All\_PGEigengap\_Unweighted & 0.03012 & 0.02725 & 0.09661 \\ 
		All\_Sum\_FAMean & 397.68878 & 360.50850 & 0.00015 & $*$ \\ 
		All\_Sum\_FiberLengthMean & 30670.09535 & 28478.19852 & 0.03582 \\ 
		All\_Sum\_FiberN & 12375.61538 & 11458.13333 & 0.00000 & $*$ \\ 
		All\_Sum\_FiberNDivLength & 548.61301 & 510.71378 & 0.00008 & $*$ \\ 
		All\_Sum\_Unweighted & 1020.80769 & 972.86667 & 0.00026 & $*$ \\ 
		Left\_AdjLMaxDivD\_FAMean & 1.37823 & 1.39812 & 0.12792 \\ 
		Left\_AdjLMaxDivD\_FiberLengthMean & 1.43638 & 1.42179 & 0.36739 \\ 
		Left\_AdjLMaxDivD\_FiberN & 1.84672 & 1.92762 & 0.12247 \\ 
		Left\_AdjLMaxDivD\_FiberNDivLength & 1.77313 & 1.80979 & 0.33521 \\ 
		Left\_AdjLMaxDivD\_Unweighted & 1.26380 & 1.25501 & 0.16858 \\ 
		Left\_HoffmanBound\_FAMean & 4.57539 & 4.44885 & 0.01512 \\ 
		Left\_HoffmanBound\_FiberLengthMean & 3.23550 & 3.25088 & 0.77158 \\ 
		Left\_HoffmanBound\_FiberN & 2.80373 & 2.74220 & 0.14090 \\ 
		Left\_HoffmanBound\_FiberNDivLength & 2.70077 & 2.64308 & 0.21782 \\ 
		Left\_HoffmanBound\_Unweighted & 4.75280 & 4.61941 & 0.00848 & $*$ \\ 
		Left\_LogSpanningForestN\_FAMean & 96.11000 & 89.25516 & 0.00106 & $*$ \\ 
		Left\_LogSpanningForestN\_FiberLengthMean & 373.09476 & 368.65582 & 0.08843 \\ 
		Left\_LogSpanningForestN\_FiberN & 300.77613 & 295.83044 & 0.00014 & $*$ \\ 
		Left\_LogSpanningForestN\_FiberNDivLength & 105.01323 & 100.80980 & 0.00288 & $*$ \\ 
		Left\_LogSpanningForestN\_Unweighted & 162.01302 & 158.88026 & 0.01336 \\ 
		Left\_MinCutBalDivSum\_FAMean & 0.00873 & 0.00273 & 0.05683 \\ 
		Left\_MinCutBalDivSum\_FiberLengthMean & 0.19822 & 0.17378 & 0.00892 & $*$ \\ 
		Left\_MinCutBalDivSum\_FiberN & 0.12848 & 0.10467 & 0.00001 & $*$ \\ 
		Left\_MinCutBalDivSum\_FiberNDivLength & 0.06926 & 0.05546 & 0.00019 & $*$ \\ 
		Left\_MinCutBalDivSum\_Unweighted & 0.19535 & 0.17339 & 0.00265 & $*$ \\ 
		Left\_MinSpanningForest\_FAMean & 14.57467 & 13.88500 & 0.06189 \\ 
		Left\_MinSpanningForest\_FiberLengthMean & 828.34729 & 834.54850 & 0.36946 \\ 
		Left\_MinSpanningForest\_FiberN & 69.30769 & 72.20000 & 0.02902 \\ 
		Left\_MinSpanningForest\_FiberNDivLength & 2.16989 & 2.25626 & 0.53695 \\ 
		Left\_MinVertexCoverBinary\_Unweighted & 48.76923 & 48.86667 & 0.69355 \\ 
		Left\_MinVertexCover\_FAMean & 14.65360 & 14.09857 & 0.01273 \\ 
		Left\_MinVertexCover\_FiberLengthMean & 1700.29684 & 1637.18742 & 0.30481 \\ 
		Left\_MinVertexCover\_FiberN & 1169.82692 & 1125.20000 & 0.06266 \\ 
		Left\_MinVertexCover\_FiberNDivLength & 58.76113 & 56.23736 & 0.06303 \\ 
		Left\_MinVertexCover\_Unweighted & 32.28846 & 32.30000 & 0.88865 \\ 
		Left\_PGEigengap\_FAMean & 0.22611 & 0.19656 & 0.00995 & $*$ \\ 
		Left\_PGEigengap\_FiberLengthMean & 0.23241 & 0.20065 & 0.02197 \\ 
		Left\_PGEigengap\_FiberN & 0.12346 & 0.10569 & 0.00382 & $*$ \\ 
		Left\_PGEigengap\_FiberNDivLength & 0.09689 & 0.08572 & 0.00223 & $*$ \\ 
		Left\_PGEigengap\_Unweighted & 0.20204 & 0.17516 & 0.01081 \\ 
		Left\_Sum\_FAMean & 197.41850 & 178.80563 & 0.00032 & $*$ \\ 
		Left\_Sum\_FiberLengthMean & 16079.40944 & 14931.40760 & 0.07487 \\ 
		Left\_Sum\_FiberN & 6071.96154 & 5641.93333 & 0.00000 & $*$ \\ 
		Left\_Sum\_FiberNDivLength & 269.09760 & 251.40080 & 0.00100 & $*$ \\ 
		Left\_Sum\_Unweighted & 519.53846 & 492.86667 & 0.00232 & $*$ \\ 
		Right\_AdjLMaxDivD\_FAMean & 1.35746 & 1.36837 & 0.36353 \\ 
		Right\_AdjLMaxDivD\_FiberLengthMean & 1.42015 & 1.41129 & 0.54264 \\ 
		Right\_AdjLMaxDivD\_FiberN & 2.05564 & 2.19134 & 0.01338 \\ 
		Right\_AdjLMaxDivD\_FiberNDivLength & 1.82146 & 1.86716 & 0.20816 \\ 
		Right\_AdjLMaxDivD\_Unweighted & 1.26684 & 1.25522 & 0.12057 \\ 
		Right\_HoffmanBound\_FAMean & 4.37886 & 4.29574 & 0.20294 \\ 
		Right\_HoffmanBound\_FiberLengthMean & 3.32686 & 3.36662 & 0.49418 \\ 
		Right\_HoffmanBound\_FiberN & 2.66511 & 2.56838 & 0.01727 \\ 
		Right\_HoffmanBound\_FiberNDivLength & 2.68679 & 2.59830 & 0.01992 \\ 
		Right\_HoffmanBound\_Unweighted & 4.60861 & 4.51407 & 0.08448 \\ 
		Right\_LogSpanningForestN\_FAMean & 93.41904 & 87.28295 & 0.00143 & $*$ \\ 
		Right\_LogSpanningForestN\_FiberLengthMean & 358.00491 & 354.73456 & 0.14280 \\ 
		Right\_LogSpanningForestN\_FiberN & 291.08563 & 285.72242 & 0.00045 & $*$ \\ 
		Right\_LogSpanningForestN\_FiberNDivLength & 100.74383 & 96.22891 & 0.00051 & $*$ \\ 
		Right\_LogSpanningForestN\_Unweighted & 154.36558 & 151.96595 & 0.01158 \\ 
		Right\_MinCutBalDivSum\_FAMean & 0.02361 & 0.01005 & 0.00807 & $*$ \\ 
		Right\_MinCutBalDivSum\_FiberLengthMean & 0.20000 & 0.17303 & 0.00768 & $*$ \\ 
		Right\_MinCutBalDivSum\_FiberN & 0.11452 & 0.10111 & 0.00563 & $*$ \\ 
		Right\_MinCutBalDivSum\_FiberNDivLength & 0.06865 & 0.06326 & 0.09375 \\ 
		Right\_MinCutBalDivSum\_Unweighted & 0.19180 & 0.16911 & 0.00492 & $*$ \\ 
		Right\_MinSpanningForest\_FAMean & 15.61479 & 14.88977 & 0.06537 \\ 
		Right\_MinSpanningForest\_FiberLengthMean & 808.14079 & 824.37649 & 0.03729 \\ 
		Right\_MinSpanningForest\_FiberN & 70.46154 & 68.93333 & 0.07096 \\ 
		Right\_MinSpanningForest\_FiberNDivLength & 2.32813 & 2.26810 & 0.46298 \\ 
		Right\_MinVertexCoverBinary\_Unweighted & 47.34615 & 47.00000 & 0.29760 \\ 
		Right\_MinVertexCover\_FAMean & 14.70648 & 14.40974 & 0.13709 \\ 
		Right\_MinVertexCover\_FiberLengthMean & 1516.99670 & 1461.52391 & 0.23679 \\ 
		Right\_MinVertexCover\_FiberN & 1175.50000 & 1166.36667 & 0.68666 \\ 
		Right\_MinVertexCover\_FiberNDivLength & 59.59421 & 58.78162 & 0.47843 \\ 
		Right\_MinVertexCover\_Unweighted & 31.61538 & 31.73333 & 0.20363 \\ 
		Right\_PGEigengap\_FAMean & 0.22838 & 0.19627 & 0.00296 & $*$ \\ 
		Right\_PGEigengap\_FiberLengthMean & 0.23840 & 0.19868 & 0.01013 \\ 
		Right\_PGEigengap\_FiberN & 0.12500 & 0.11049 & 0.00869 & $*$ \\ 
		Right\_PGEigengap\_FiberNDivLength & 0.10075 & 0.09371 & 0.03033 \\ 
		Right\_PGEigengap\_Unweighted & 0.20584 & 0.17429 & 0.00242 & $*$ \\ 
		Right\_Sum\_FAMean & 190.48228 & 172.48988 & 0.00062 & $*$ \\ 
		Right\_Sum\_FiberLengthMean & 13952.01182 & 13003.32443 & 0.04620 \\ 
		Right\_Sum\_FiberN & 5935.73077 & 5525.26667 & 0.00001 & $*$ \\ 
		Right\_Sum\_FiberNDivLength & 262.31420 & 246.32048 & 0.00180 & $*$ \\ 
		Right\_Sum\_Unweighted & 477.38462 & 454.86667 & 0.00068 & $*$ \\ 
		\end{longtable}
	
	\subsection*{Scale 234, round 1}
	\begin{longtable}{l | cccc}
		Property & Female & Male & p-value &  \\ 
		All\_AdjLMaxDivD\_FAMean & 2.15050 & 2.14489 & 0.86385 \\ 
		All\_AdjLMaxDivD\_FiberLengthMean & 2.35868 & 2.34695 & 0.80876 \\ 
		All\_AdjLMaxDivD\_FiberN & 5.14838 & 5.00652 & 0.35870 \\ 
		All\_AdjLMaxDivD\_FiberNDivLength & 5.17072 & 4.78287 & 0.02543 \\ 
		All\_AdjLMaxDivD\_Unweighted & 1.89062 & 1.84578 & 0.06482 \\ 
		All\_HoffmanBound\_FAMean & 3.63940 & 3.62013 & 0.57408 \\ 
		All\_HoffmanBound\_FiberLengthMean & 2.92490 & 2.98466 & 0.17340 \\ 
		All\_HoffmanBound\_FiberN & 2.23619 & 2.26557 & 0.30055 \\ 
		All\_HoffmanBound\_FiberNDivLength & 2.20178 & 2.23871 & 0.13550 \\ 
		All\_HoffmanBound\_Unweighted & 3.73661 & 3.72935 & 0.82472 \\ 
		All\_LogSpanningForestN\_FAMean & 446.86116 & 416.54482 & 0.03232 \\ 
		All\_LogSpanningForestN\_FiberLengthMean & 2324.68381 & 2325.52712 & 0.96824 \\ 
		All\_LogSpanningForestN\_FiberN & 1456.24015 & 1445.53700 & 0.36683 \\ 
		All\_LogSpanningForestN\_FiberNDivLength & 149.01647 & 138.16817 & 0.15229 \\ 
		All\_LogSpanningForestN\_Unweighted & 942.01654 & 944.27877 & 0.83734 \\ 
		All\_MinCutBalDivSum\_FAMean & 0.00000 & 0.00000 & nan \\ 
		All\_MinCutBalDivSum\_FiberLengthMean & 0.00769 & 0.00723 & 0.57442 \\ 
		All\_MinCutBalDivSum\_FiberN & 0.02405 & 0.02168 & 0.21132 \\ 
		All\_MinCutBalDivSum\_FiberNDivLength & 0.00000 & 0.00000 & 0.45008 \\ 
		All\_MinCutBalDivSum\_Unweighted & 0.00898 & 0.00834 & 0.32475 \\ 
		All\_MinSpanningForest\_FAMean & 98.19730 & 92.47667 & 0.00151 & $*$ \\ 
		All\_MinSpanningForest\_FiberLengthMean & 5358.83904 & 5379.38212 & 0.44199 \\ 
		All\_MinSpanningForest\_FiberN & 481.46154 & 479.20000 & 0.45787 \\ 
		All\_MinSpanningForest\_FiberNDivLength & 18.53246 & 18.36575 & 0.71037 \\ 
		All\_MinVertexCoverBinary\_Unweighted & 276.15385 & 280.33333 & 0.12225 \\ 
		All\_MinVertexCover\_FAMean & 89.53747 & 87.25805 & 0.06974 \\ 
		All\_MinVertexCover\_FiberLengthMean & 8136.04292 & 7957.20990 & 0.48358 \\ 
		All\_MinVertexCover\_FiberN & 2430.61538 & 2344.50000 & 0.00056 & $*$ \\ 
		All\_MinVertexCover\_FiberNDivLength & 129.82332 & 126.64639 & 0.02087 \\ 
		All\_MinVertexCover\_Unweighted & 222.57692 & 223.33333 & 0.39844 \\ 
		All\_PGEigengap\_FAMean & 0.01106 & 0.01201 & 0.54543 \\ 
		All\_PGEigengap\_FiberLengthMean & 0.00860 & 0.00960 & 0.45409 \\ 
		All\_PGEigengap\_FiberN & 0.01894 & 0.01927 & 0.89543 \\ 
		All\_PGEigengap\_FiberNDivLength & 0.01773 & 0.01767 & 0.97772 \\ 
		All\_PGEigengap\_Unweighted & 0.00995 & 0.01067 & 0.59117 \\ 
		All\_Sum\_FAMean & 1033.36931 & 961.08503 & 0.00297 & $*$\\ 
		All\_Sum\_FiberLengthMean & 74747.99556 & 71461.78993 & 0.18467 \\ 
		All\_Sum\_FiberN & 13609.34615 & 12823.40000 & 0.00000 & $*$\\ 
		All\_Sum\_FiberNDivLength & 652.17760 & 623.38731 & 0.00139 & $*$\\ 
		All\_Sum\_Unweighted & 2801.69231 & 2746.20000 & 0.21290 \\ 
		Left\_AdjLMaxDivD\_FAMean & 2.14627 & 2.14335 & 0.93401 \\ 
		Left\_AdjLMaxDivD\_FiberLengthMean & 2.29338 & 2.29214 & 0.97718 \\ 
		Left\_AdjLMaxDivD\_FiberN & 4.03186 & 4.16381 & 0.29128 \\ 
		Left\_AdjLMaxDivD\_FiberNDivLength & 3.93717 & 3.84897 & 0.38654 \\ 
		Left\_AdjLMaxDivD\_Unweighted & 1.86339 & 1.81508 & 0.04174 \\ 
		Left\_HoffmanBound\_FAMean & 3.74670 & 3.77335 & 0.55549 \\ 
		Left\_HoffmanBound\_FiberLengthMean & 2.94312 & 2.99233 & 0.25660 \\ 
		Left\_HoffmanBound\_FiberN & 2.51168 & 2.47461 & 0.28318 \\ 
		Left\_HoffmanBound\_FiberNDivLength & 2.44470 & 2.45140 & 0.85286 \\ 
		Left\_HoffmanBound\_Unweighted & 3.82814 & 3.84621 & 0.65499 \\ 
		Left\_LogSpanningForestN\_FAMean & 212.18613 & 197.08273 & 0.04326 \\ 
		Left\_LogSpanningForestN\_FiberLengthMean & 1159.44274 & 1165.33847 & 0.58696 \\ 
		Left\_LogSpanningForestN\_FiberN & 723.10349 & 723.01322 & 0.98899 \\ 
		Left\_LogSpanningForestN\_FiberNDivLength & 70.44766 & 65.77187 & 0.31060 \\ 
		Left\_LogSpanningForestN\_Unweighted & 467.24325 & 470.94213 & 0.52729 \\ 
		Left\_MinCutBalDivSum\_FAMean & 0.00000 & 0.00000 & nan \\ 
		Left\_MinCutBalDivSum\_FiberLengthMean & 0.09355 & 0.07667 & 0.00655 & $*$\\ 
		Left\_MinCutBalDivSum\_FiberN & 0.07158 & 0.05914 & 0.00062 & $*$\\ 
		Left\_MinCutBalDivSum\_FiberNDivLength & 0.00000 & 0.00000 & nan \\ 
		Left\_MinCutBalDivSum\_Unweighted & 0.09416 & 0.07896 & 0.00153 & $*$\\ 
		Left\_MinSpanningForest\_FAMean & 47.28302 & 44.78250 & 0.00239 & $*$\\ 
		Left\_MinSpanningForest\_FiberLengthMean & 2702.23206 & 2712.65026 & 0.49327 \\ 
		Left\_MinSpanningForest\_FiberN & 244.11538 & 244.46667 & 0.89014 \\ 
		Left\_MinSpanningForest\_FiberNDivLength & 9.45842 & 9.50259 & 0.88229 \\ 
		Left\_MinVertexCoverBinary\_Unweighted & 137.19231 & 140.00000 & 0.06105 \\ 
		Left\_MinVertexCover\_FAMean & 43.50481 & 42.59720 & 0.16942 \\ 
		Left\_MinVertexCover\_FiberLengthMean & 4136.87086 & 4052.71473 & 0.55895 \\ 
		Left\_MinVertexCover\_FiberN & 1168.19231 & 1153.66667 & 0.46021 \\ 
		Left\_MinVertexCover\_FiberNDivLength & 63.94002 & 64.04107 & 0.92511 \\ 
		Left\_MinVertexCover\_Unweighted & 111.38462 & 112.26667 & 0.09259 \\ 
		Left\_PGEigengap\_FAMean & 0.08402 & 0.07554 & 0.28777 \\ 
		Left\_PGEigengap\_FiberLengthMean & 0.08669 & 0.07722 & 0.29463 \\ 
		Left\_PGEigengap\_FiberN & 0.06812 & 0.05737 & 0.09675 \\ 
		Left\_PGEigengap\_FiberNDivLength & 0.05084 & 0.04481 & 0.18106 \\ 
		Left\_PGEigengap\_Unweighted & 0.07190 & 0.06398 & 0.24844 \\ 
		Left\_Sum\_FAMean & 504.02280 & 470.30921 & 0.01077 \\ 
		Left\_Sum\_FiberLengthMean & 38178.70022 & 36255.83071 & 0.19037 \\ 
		Left\_Sum\_FiberN & 6716.53846 & 6389.20000 & 0.00107 & $*$\\ 
		Left\_Sum\_FiberNDivLength & 322.55630 & 311.23280 & 0.04079 \\ 
		Left\_Sum\_Unweighted & 1401.80769 & 1380.33333 & 0.39428 \\ 
		Right\_AdjLMaxDivD\_FAMean & 2.00996 & 2.02718 & 0.61502 \\ 
		Right\_AdjLMaxDivD\_FiberLengthMean & 2.15381 & 2.18170 & 0.41400 \\ 
		Right\_AdjLMaxDivD\_FiberN & 4.11898 & 4.41926 & 0.03397 \\ 
		Right\_AdjLMaxDivD\_FiberNDivLength & 3.79534 & 3.75488 & 0.70781 \\ 
		Right\_AdjLMaxDivD\_Unweighted & 1.79189 & 1.77141 & 0.38704 \\ 
		Right\_HoffmanBound\_FAMean & 3.63008 & 3.59884 & 0.45778 \\ 
		Right\_HoffmanBound\_FiberLengthMean & 3.00591 & 3.02300 & 0.69490 \\ 
		Right\_HoffmanBound\_FiberN & 2.40837 & 2.33314 & 0.00150 & $*$\\ 
		Right\_HoffmanBound\_FiberNDivLength & 2.45857 & 2.38848 & 0.01602 \\ 
		Right\_HoffmanBound\_Unweighted & 3.71704 & 3.69299 & 0.50645 \\ 
		Right\_LogSpanningForestN\_FAMean & 228.90719 & 215.28259 & 0.07936 \\ 
		Right\_LogSpanningForestN\_FiberLengthMean & 1154.04516 & 1148.91122 & 0.63377 \\ 
		Right\_LogSpanningForestN\_FiberN & 724.05083 & 716.03208 & 0.22608 \\ 
		Right\_LogSpanningForestN\_FiberNDivLength & 72.92465 & 68.45678 & 0.30478 \\ 
		Right\_LogSpanningForestN\_Unweighted & 467.61765 & 466.56728 & 0.85195 \\ 
		Right\_MinCutBalDivSum\_FAMean & 0.00050 & 0.00000 & 0.19303 \\ 
		Right\_MinCutBalDivSum\_FiberLengthMean & 0.10021 & 0.08439 & 0.01271 \\ 
		Right\_MinCutBalDivSum\_FiberN & 0.07599 & 0.06701 & 0.00641 & $*$\\ 
		Right\_MinCutBalDivSum\_FiberNDivLength & 0.00034 & 0.00000 & 0.18042 \\ 
		Right\_MinCutBalDivSum\_Unweighted & 0.09573 & 0.08171 & 0.01034 \\ 
		Right\_MinSpanningForest\_FAMean & 50.98056 & 47.79220 & 0.00435 & $*$\\ 
		Right\_MinSpanningForest\_FiberLengthMean & 2655.83115 & 2655.71544 & 0.99483 \\ 
		Right\_MinSpanningForest\_FiberN & 238.96154 & 236.00000 & 0.15420 \\ 
		Right\_MinSpanningForest\_FiberNDivLength & 9.28191 & 9.00082 & 0.18645 \\ 
		Right\_MinVertexCoverBinary\_Unweighted & 138.30769 & 140.00000 & 0.25603 \\ 
		Right\_MinVertexCover\_FAMean & 45.80119 & 44.57707 & 0.07765 \\ 
		Right\_MinVertexCover\_FiberLengthMean & 3994.00115 & 3884.90036 & 0.36802 \\ 
		Right\_MinVertexCover\_FiberN & 1144.80769 & 1129.73333 & 0.41752 \\ 
		Right\_MinVertexCover\_FiberNDivLength & 62.35579 & 61.50301 & 0.47854 \\ 
		Right\_MinVertexCover\_Unweighted & 111.09615 & 111.10000 & 0.99385 \\ 
		Right\_PGEigengap\_FAMean & 0.08312 & 0.07683 & 0.33378 \\ 
		Right\_PGEigengap\_FiberLengthMean & 0.08538 & 0.07887 & 0.40909 \\ 
		Right\_PGEigengap\_FiberN & 0.06631 & 0.06080 & 0.28067 \\ 
		Right\_PGEigengap\_FiberNDivLength & 0.05084 & 0.04854 & 0.52890 \\ 
		Right\_PGEigengap\_Unweighted & 0.07102 & 0.06430 & 0.25554 \\ 
		Right\_Sum\_FAMean & 517.36095 & 481.68012 & 0.00745 & $*$\\ 
		Right\_Sum\_FiberLengthMean & 35857.03890 & 34486.76733 & 0.26347 \\ 
		Right\_Sum\_FiberN & 6524.53846 & 6187.46667 & 0.00050 & $*$\\ 
		Right\_Sum\_FiberNDivLength & 312.50248 & 299.09835 & 0.01170 \\ 
		Right\_Sum\_Unweighted & 1368.00000 & 1339.06667 & 0.20464 \\ 
		\end{longtable}

}

\end{document}